\documentclass[prb,twocolumn,showpacs]{revtex4-1}
\usepackage{amsfonts}
\usepackage{amsmath}
\usepackage{amssymb}
\usepackage{graphicx}
\usepackage{epstopdf}

\begin{document}

\title{Duality in topological superconductors and topological ferromagnetic insulators in a honeycomb lattice}
\author{Shin-Ming Huang$^{1}$, Wei-Feng Tsai$^{2}$, Chung-Hou Chung$^{3,4}$%
, and Chung-Yu Mou$^{1,4,5}$}

\affiliation{$^{1}$ Department of Physics, National Tsing Hua University, Hsinchu 30043,
Taiwan, 300, R.O.C.}
\affiliation{$^{2}$Department of Physics, National Sun Yat-Sen
University, Kaohsiung, Taiwan, R.O.C.}
\affiliation{$^{3}$Electrophysics
Department, National Chiao-Tung University, Hsinchu, Taiwan, R.O.C.}
\affiliation{$^{4}$Physics Division, National Center for Theoretical Sciences, P.O.Box
2-131, Hsinchu, Taiwan, R.O.C.}
\affiliation{$^{5}$Department of Physics, National Tsing Hua University, Hsinchu 30043,
Taiwan, 300, R.O.C.} \affiliation{$^{7}$Institute of Physics, Academia
Sinica, Nankang, Taiwan, R.O.C.}

\begin{abstract}
The ground state of large Hubbard $U$ limit of a honeycomb lattice near
half-filling is known to be a singlet $d+id$-wave superconductor. It is also
known that this $d+id$ superconductor exhibits a chiral $p+ip$ pairing
locally at the Dirac cone, characterized by a $2\mathbb{Z}$ topological
invariant. By constructing a dual transformation, we demonstrate that this $2\mathbb{Z}$ 
topological superconductor is equivalent to a collection of two topological ferromagnetic insulators. 
As a result of the duality, the topology of
the electronic structures for a $d+id$ superconductor is controllable via
the change of the chemical potential by tuning the gate voltage. In
particular, instead of being always a chiral superconductor, we find that
the $d+id$ superconductor undergoes a topological phase transition from a chiral
superconductor to a quasi-helical superconductor as the gap amplitude or the
chemical potential decreases. The quasi-helical superconducting phase is found to
be characterized by a topological invariant in the pseudo-spin charge
sector with vanishing both the Chern number and the spin Chern number. We further elucidate the
topological phase transition by analyzing the relationship between the
topological invariant and the rotation symmetry. Due to the angular momentum
carried by the gap function and spin-orbit interactions, we show that by
placing $d+id$ superconductors in proximity to ferromagnets, varieties of
chiral superconducting phases characterized by higher Chern numbers can be
accessed, providing a new platform for hosting large numbers of Majorana
modes at edges.
\end{abstract}


\date{\today }
\maketitle

\section{Introduction}

A $\mathbb{Z}$ topological insulator in two dimensions, also called a Chern insulator, characterized by the
Chern number, is an electronic system with broken time-reversal symmetry
(TRS), showing a quantized Hall conductivity~\cite{TKNN} and protected
gapless edge modes.~\cite{Halperin1982,Hatsugai1993} An example is the
quantum anomalous Hall (QAH) effect, being achieved by a magnetic exchange
field. \cite{CZChang2013, Huang2014} After the discovery of the $\mathbb{Z}_{2}$ quantum
spin Hall (QSH) insulator,~\cite%
{Kane2005a,Kane2005b,Bernevig2006,Konig2007,Roth2009,Zhang2009} it is
realized that symmetries play important roles in classifying a topological state. According to the AZ-classification scheme by TRS, particle-hole
symmetry (PHS), and chiral symmetry, \cite{AZa,AZb} topological insulators
and superconductors in different non-spatial symmetry classes belong to
different category and are not connected in topology.~\cite%
{Schnyder2008,Kitaev2009,Chiu2015} From symmetry point of view on the energy
spectrum, however, superconductors have energy gaps and can be considered as
an insulator with conduction and valence bands related by the particle-hole
symmetry. It is therefore interesting to explore possible connections of
superconductors and insulators in the topology of electronic structures.
Such connection can be useful in help searching possible realizations of
Majorana modes in superconductors.

In the past, the investigation on topological superconductors had been
mostly focusing on the triplet pairing superconductors with TRS breaking.
These topological superconductors host chiral Majorana fermions at edges but
there are very few confirmed observations.~\cite{Lee1997,Tou1996,Eremin2004,Maeno2012} Until recently, thanks to the seminal
work by Fu and Kane,\cite{Fu2008,Mourik2012} it is now realized that the
proximity of a topological insulator to a singlet \textit{s}-wave
superconductor provides an alternative way to topological superconductors. The
combined effect of $s$-wave and the spin-orbit coupling (SOC) plays an
important role in hosting Majorana fermions. In this case, the most general
topological superconductors are time-reversal invariant helical superconductors (HSCs), characterized by a $\mathbb{Z}_{2}$ topological invariant.~\cite{Oreg2010,FZhang2013a,Kao2015} More recently, possible solutions of
spin-singlet topological superconductors via spontaneously TRS breaking are
proposed.~\cite{Laughlin1998,Moore2004,Rosenstein2013,CCLee2009} In
particular, the ground state for electrons on the honeycomb lattice in the
large Hubbard $U$ limit are shown to be of chiral $d$-wave pairing symmetry, $
d_{x^{2}-y^{2}}+id_{xy}$ ($d+id$),~\cite{chirald} which indicates that the
honeycomb lattice might be a good platform to host chiral Majorana modes.

In this paper, we explore the topology of electronic structures for $d+id$
-wave superconductivity in a honeycomb lattice. We first demonstrate that the
electronic structures of a $d+id$ superconductor and a ferromagnetic insulator on a honeycomb lattice are interchanged under a dual
transformation. As a result of the duality, the topology of the electronic
structures for a $d+id$ superconductor is equivalent to that of a
ferromagnetic insulator. The idea of a dual transformation from an $s$-wave superconductor with the Rashba SOC to a $p$-wave superconductor was proposed before,~\cite{MSato2009} but the dual system was obscure in physical meaning. The duality in our models is between two realistic ones and indicates equivalent topological structures underlying. In the ferromagnetic insulator, QAH and spin Chern insulating (SCI)/QSH phases can be found. QAH and SCI/QSH phases are characterized by a non-vanishing Chern number and a spin Chern number, respectively. The spin Chern number is an alternative way to characterize the $\mathbb{Z}_{2}$ QSH insulator,~\cite{Sheng2006a} which originates from the intuition that when the fiber bundles of filled states are projected into spin-up and spin-down sectors the nontrivial topological structure can be found in each spin section.~\citep{Prodan2009} The spin Chern number is confirmed to be a robust topological invariant against disorder or spin-nonconserving interactions (such as Rashba SOC). Unlike the $\mathbb{Z}_{2}$ invariant, the spin Chern number can be generalized to the case when TRS is broken as long as the band gap remains open to preserve a finite spin polarization.~\citep{Sheng2006a,Prodan2009,Yang2011,Sheng2013} With the aid of the duality, a phase analogous to SCI/QSH, termed as quasi-HSC, is found in $d+id$ superconductors. Here "quasi" indicates that there is no exact time-reversal partner between counter-propagating edge modes. The quasi-HSC phase has no spin Chern number and is characterized by a pseudo-spin Chern number in the charge sector not in the spin sector. As the topological phase undergoes a transition to a chiral superconductor (CSC) for larger gap amplitudes or the
chemical potentials, its topology is described by the Chern number and it is realized as a combination of two QAH systems with a charge conjugate relation. Furthermore, We
elucidate the topological phase by analyzing the relationship
between the topological invariant and the rotation symmetry associated with
the angular momentum carried by the gap function and SOC. When the $d+id$ superconductor is in proximity to a
ferromagnet, the superconducting state coexists with ferromagnetism. We find
that varieties of chiral superconducting phases characterized by higher
Chern numbers can be accessed, which provides a new platform for hosting
large numbers of Majorana modes at edges.

\section{Theoretical Models}

In this section, we start by considering the topological ferromagnetic insulator in Sec. \ref{sec:theory-TI}. The phase diagram will be
constructed. The spin-singlet $d+id$-wave topological superconductor will be
investigate in Sec. \ref{sec:theory-classD1}. The duality relation with the topological ferromagnetic insulator will be clarified. Finally, in Sec. \ref%
{sec:theory-classD2}, we combine both models by considering the situation
when the $d+id$ superconductor is in proximity to a ferromagnet. The
topological phase diagram for the case when the superconductivity coexists
with ferromagnetism is constructed. The effect of duality on the phase
diagram and the relation to the rotation symmetry will be given.

\subsection{Class A insulator}

\label{sec:theory-TI} 
\begin{figure}[tbp]
\includegraphics[width=0.48\textwidth]{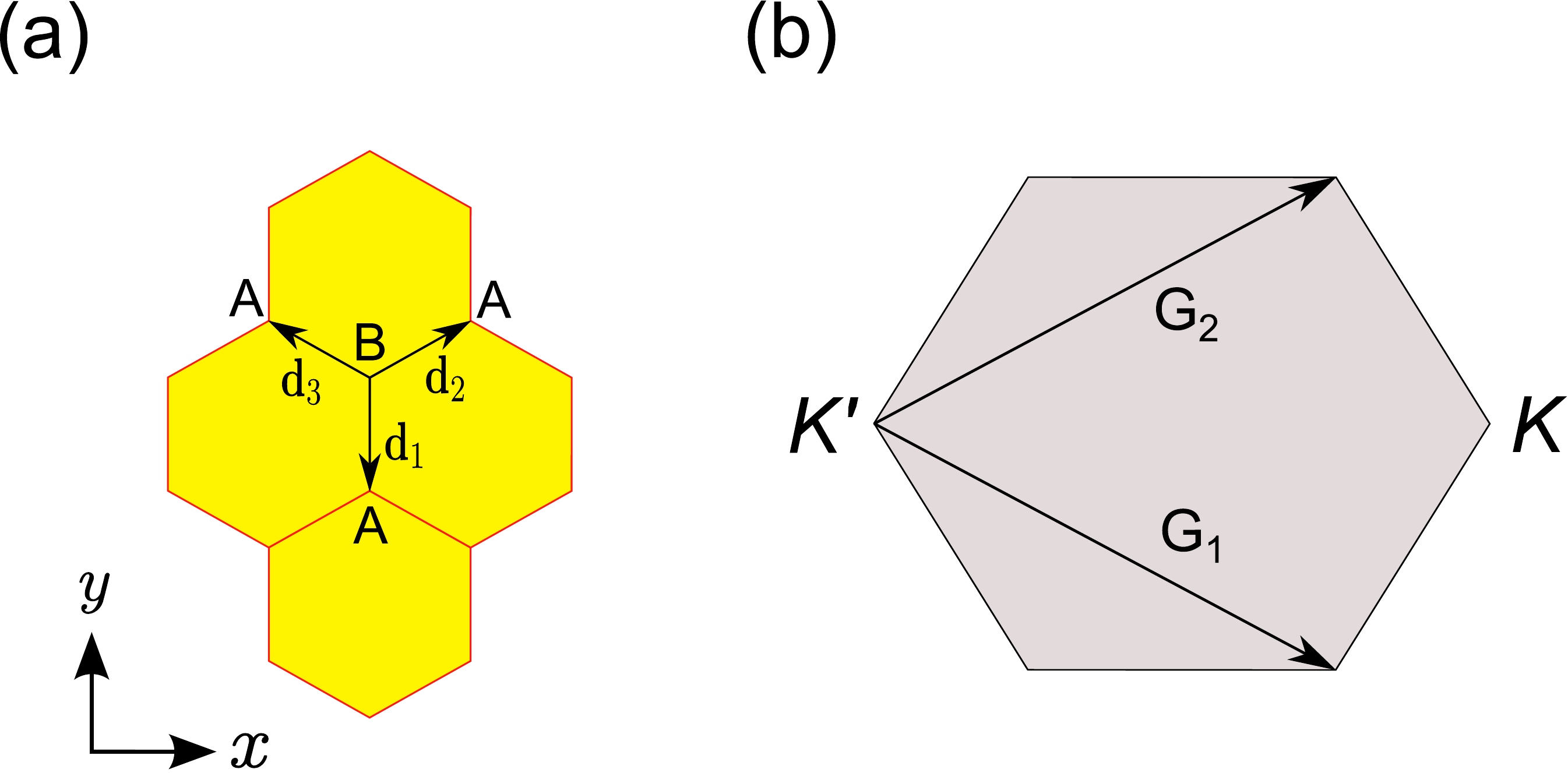}\newline
\caption{(Color online) (a) Three nearest neighbour vectors $\mathbf{d}_{1,2,3}$ and
coordinates in a honeycomb lattice. (b) Brillouin zone and reciprocal
lattice vectors, $\mathbf{G}_{1}$ and $\mathbf{G}_{2}$}
\label{fig:d_vectors}
\end{figure}

We start by considering the Kane-Mele model in the
presence of the exchange field $-M$
\begin{eqnarray}
\hat{H}_{\mathrm{FM}} &=&-t\sum_{\left\langle i,j\right\rangle }c_{i}^{\dag
}c_{j}+i\frac{\lambda _{\mathrm{SO}}}{3\sqrt{3}}\sum_{\left\langle
\left\langle i,j\right\rangle \right\rangle }\nu _{ij}c_{i}^{\dag }\sigma
_{z}c_{j}  \label{HFM0} \\
&&+i\frac{2\lambda _{\mathrm{R}}}{3}\sum_{\left\langle i,j\right\rangle
}c_{i}^{\dag }\hat{z}\cdot \left( \mathbf{\sigma \times \hat{d}}_{ij}\right)
c_{j}-M\sum_{i}c_{i}^{\dag }\sigma _{z}c_{i}.  \notag
\end{eqnarray}%
Here the first term describes the hopping, the second term is the intrinsic
SOC, the third one is the Rashba SOC, and the last term is the exchange
field. $c_{i}^{\dag }=\left( c_{i\uparrow }^{\dag },c_{i\downarrow }^{\dag
}\right) $ is the electron creation operator on site \textit{i}, $\mathbf{%
\sigma }$ is the Pauli matrix, and $\left\langle i,j\right\rangle $ and $%
\left\langle \left\langle i,j\right\rangle \right\rangle $ denote \textit{i}
and \textit{j} being nearest-neighbor (NN) and next-nearest-neighbor (NNN)
sites, respectively. $\mathbf{\hat{d}}_{ij}=\mathbf{d}_{ij}/\left\vert
\mathbf{d}_{ij}\right\vert $ are unit vectors connecting site \textit{j} and
\textit{i}. Three NN vectors $\mathbf{d}_{l}$ ($l=1,2,3$) along with the
coordinates are shown in Fig. \ref{fig:d_vectors}. $\nu _{ij}=\mathrm{sgn}(%
\hat{z}\cdot \mathbf{d}_{kj}\times \mathbf{d}_{ik})=\pm 1$ for \textit{ij}
being connected by $\mathbf{d}_{kj}$ and $\mathbf{d}_{ik}$.

\begin{figure}[tbp]
\includegraphics[width=0.49\textwidth]{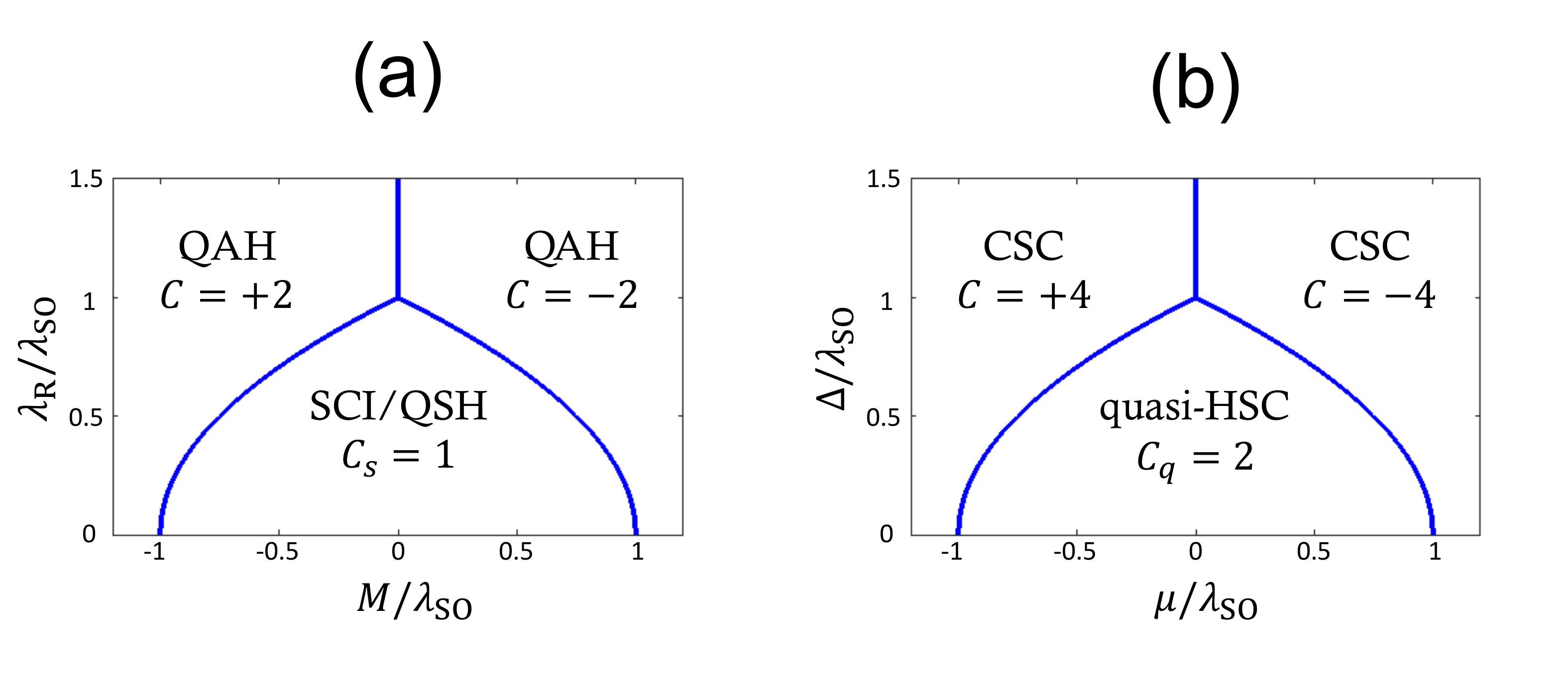}\newline
\caption{(Color online) (a) Phase diagram for the ferromagnetic system in Eq. (%
\protect\ref{HFM0}). QAH and SCI/QSH denote phases of quantum anomalous Hall insulator and
spin Chern insulator/quantum spin Hall insulator, which are dubbed by nonzero Chern number $C$ and spin Chern number $C_{s}=(C_{\uparrow }-C_{\downarrow })/2 $, respectively. (b) The phase diagram for the
superconducting state in Eq. (\protect\ref{Eq14}). Similar to (a) with substitutions: $%
M\rightarrow \protect\mu $, $\protect\lambda _{\mathrm{R}}\rightarrow \Delta
$, doubled Chern numbers, $C=\pm 2 \rightarrow C=\pm 4$, and $C_s=1 \rightarrow C_q=2$
(see text), and CSC/quasi-HSC (chiral/quasi-helical superconductor) in place of QAH/SCI. Note that for the QAH and CSC phases, one requires $\lambda _{\mathrm{R}} \neq 0$ and $\Delta \neq 0$, respectively. }
\label{fig:PD_FM}
\end{figure}

After the Fourier transformation, the Bloch Hamiltonian becomes
\begin{equation}
\mathcal{H}_{\mathrm{FM}}(\mathbf{k})=\left(
\begin{array}{cccc}
\Lambda _{\mathbf{k}}-M & T_{\mathbf{k}} & 0 & R_{\mathbf{k}} \\
T_{\mathbf{k}}^{\ast } & -\Lambda _{\mathbf{k}}-M & -R_{-\mathbf{k}} & 0 \\
0 & -R_{-\mathbf{k}}^{\ast } & -\Lambda _{\mathbf{k}}+M & T_{\mathbf{k}} \\
R_{\mathbf{k}}^{\ast } & 0 & T_{\mathbf{k}}^{\ast } & \Lambda _{\mathbf{k}}+M%
\end{array}%
\right)  \label{HFM}
\end{equation}%
in the basis $c_{\mathrm{FM}}(\mathbf{k})=\left(
\begin{array}{cccc}
c_{A\mathbf{k}\uparrow } & c_{B\mathbf{k}\uparrow } & c_{A\mathbf{k}%
\downarrow } & c_{B\mathbf{k}\downarrow }%
\end{array}%
\right) ^{T}$, with
\begin{eqnarray}
T_{\mathbf{k}} &=&-t\sum_{l}e^{-i\mathbf{k\cdot d}_{l}}, \\
\Lambda _{\mathbf{k}} &=&\frac{2\lambda _{\mathrm{SO}}}{3\sqrt{3}}\sum_{l%
\text{ }(\mathbf{d}_{4}=\mathbf{d}_{1})}\sin \mathbf{k\cdot }\left( \mathbf{d%
}_{l}-\mathbf{d}_{l+1}\right) , \\
R_{\mathbf{k}} &=&-\frac{2\lambda _{\mathrm{R}}}{3}\sum_{l}e^{-i\theta
_{l}}e^{-i\mathbf{k\cdot d}_{l}},
\end{eqnarray}%
where $\theta _{l}$ is the polar angle of the vector $\mathbf{d}_{l}$. We will set the lattice constant and Planck constant as unity, $\hbar =a=1$.

The topological phases of the Kane-Mele model have been understood and are
summarized in Fig. \ref{fig:PD_FM}(a). When $M=0$ and $\lambda _{\mathrm{SO}}>|\lambda _{%
\mathrm{R}}|$ (assume $\lambda _{\mathrm{SO}}>0$), the low-energy physics is
described by a massive Dirac fermion (and its time-reversal partner with an
opposite mass) at $K$ and $K^{\prime }$ points where Berry phases $\pi $ ($-\pi $) are underlying. This is a QSH phase characterized by a $\mathbb{Z}_{2}$ topological invariant. When $M \neq 0$ and TRS is broken, the QSH phase is replaced by the SCI phase when $\lambda _{\mathrm{SO}}>0$.
In this case, even though one still gets counter-propagating edge states, these edges states are
not robust and can be gapped under perturbations.\cite{Sheng2013}
However, the spin Chern number is still well-defined\cite{Prodan2009} and is intact.~\cite{Yang2011} The robustness of the value for the spin Chern number is due to the spin gap, associated with
the band gap, that two occupied states remain carrying definite opposite
spin projections as TRS preserves. \cite{Sheng2013} The topological phase
transition happens at the closing of the band gap, given by
\begin{equation}
\left\vert M\right\vert =\left( \lambda _{\mathrm{SO}}^{2}-\lambda _{\mathrm{%
R}}^{2}\right) /\lambda _{\mathrm{SO}}.  \label{QPT}
\end{equation}%
Over the boundary in Eq. (\ref{QPT}) with $\lambda _{\mathrm{R}}$ and $M$
both finite, the system enters the QAH phase with Chern number $C=\pm 2$.~%
\cite{Qiao2010}

\begin{figure}[tbp]
\includegraphics[width=0.48\textwidth]{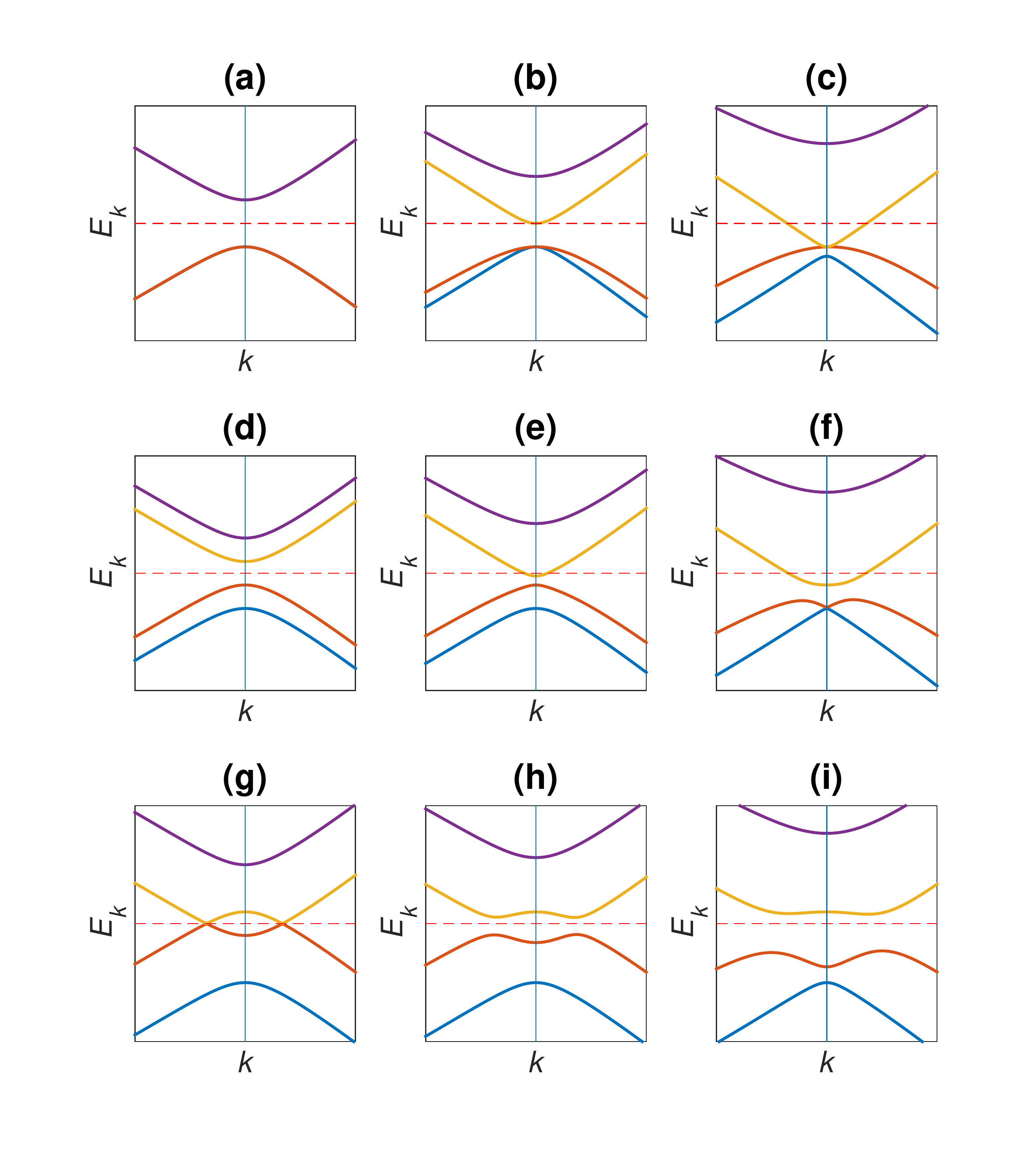}\newline
\caption{(Color online) Energy dispersion around the $K$ point (vertical
line) for the topological insulator with ferromagnetism described by Eq. (%
\protect\ref{HFM0}). Different combinations of $M$ and $\protect\lambda _{%
\mathrm{R}}$ are given. $M$ increases from top to bottom; $M/\protect\lambda %
_{\mathrm{SO}}=0$ [(a), (b), (c)], $0.5$ [(d), (e), (f)], and $1.5$ [(g),
(h), (i)], respectively. On the other hand, $\protect\lambda _{\text{R}}$
increases from left to right; $\protect\lambda _{\mathrm{R}}/\protect\lambda %
_{\mathrm{SO}}=0$ [(a), (d), (g)], $0.5$ [(b), (e), (h)], and $1.2$ [(c),
(f), (i)], respectively. The horizontal dashed line denotes zero energy as
the unbiased chemical potential. If one can shift the chemical potential
within the gap between middle two bands, topological phases are achieved:
(a) and (b) are in the QSH phase. (d) and (e) are in the SCI phase. (c) and (g) are in the critical
semimetal phase, preparing for entering the QAH phase, and (f), (h), (i) are
in the QAH phase. }
\label{fig:Ek_FM}
\end{figure}

To elucidate the topological phase transition more clearly, the band structures
near $K$ for different cases of $\lambda _{\mathrm{R}}$ and $M$ are shown in
Fig. \ref{fig:Ek_FM} with $\lambda _{\mathrm{R}}/\lambda _{\mathrm{SO}}=0$,
0.5, 1.2 from left to right columns and $M/\lambda _{\mathrm{SO}}=0$, 0.5,
1.5 from top to bottom rows. At $\lambda _{\mathrm{R}}=M=0$, Fig. \ref
{fig:Ek_FM}(a), bands are spin-degenerate and are separated by a gap of $
2\lambda _{\mathrm{SO}}$. Non-vanishing $\lambda _{\mathrm{R}}$ and/or $M$
lifts the spin degeneracy and brings the conduction and the valence bands
approaching to each other. The resulting phase is an SCI phase. 
As a result, the band gap prevents the band
inversion to occur immediately and thus protects the resulting SCI phase. Once $
\lambda _{\mathrm{R}}$ and $M$ terms overcome the gap by $\lambda _{\mathrm{%
SO}}$, due to that $\lambda _{\mathrm{R}}$ and $M$ terms anti-commute, two
bands must anti-cross and exchange the Chern number, resulting in a QAH
phase.

We now illustrate the change of the Chern number from the point of view of
exchanging rotation eigenvalues. In an insulator with point group
symmetries, the eigenvalues of symmetry operators at high-symmetry points
of the ground state determine the Chern number up to a multiple of $n$ in the presence of $n$-fold axis. By Ref. [\onlinecite{Fang2012a}], a three-fold-symmetric insulator as our system gives
\begin{equation}
e^{-i2\pi C/3}=\prod\limits_{i\in \mathrm{occ.}}\eta _{i}(\Gamma )\eta
_{i}(K)\eta _{i}(K^{\prime }),
\end{equation}%
where $\eta _{i}(\mathbf{k})$ is the eigenvalues of the three-fold rotation.
Since the band inversion happens at $K$ and $K^{\prime}$, we can study the
change of the Chern number by the change of rotation eigenvalues at $K$ and $%
K^{\prime}$. The eigenenergies of the Bloch Hamiltonian at $K$ and $%
K^{\prime}$, Eq. (\ref{HFM}), are
\begin{equation}
\begin{array}{l}
E_{1}=-\lambda _{\mathrm{SO}}-M, \\
E_{2}=-\lambda _{\mathrm{SO}}+M, \\
E_{3}=\lambda _{\mathrm{SO}}-\sqrt{M^{2}+4\lambda _{\mathrm{R}}^{2}}, \\
E_{4}=\lambda _{\mathrm{SO}}+\sqrt{M^{2}+4\lambda _{\mathrm{R}}^{2}}.%
\end{array}%
\end{equation}%
Here the rotation eigenvalues of these four states at $K$ are given by
\begin{equation}
\begin{array}{l}
\eta _{1}^{K}=e^{i\pi /3}, \\
\eta _{2}^{K}=e^{-i3\pi /3}, \\
\eta _{3}^{K}=e^{-i\pi /3}, \\
\eta _{4}^{K}=e^{-i\pi /3},%
\end{array}
\label{rot_K}
\end{equation}
respectively, while for the $K^{\prime}$ point, the rotation eigenvalues of
these four eigenenergies are
\begin{equation}
\begin{array}{l}
\eta _{1}^{K^{\prime }}=e^{i3\pi /3}, \\
\eta _{2}^{K^{\prime }}=e^{-i\pi /3}, \\
\eta _{3}^{K^{\prime }}=e^{i\pi /3}, \\
\eta _{4}^{K^{\prime }}=e^{i\pi /3},%
\end{array}
\label{rot_mK}
\end{equation}
respectively. Detailed derivation of rotation eigenvalues can be found in
Appendix \ref{sec:app-rotation_normal}. Consider the case of $M>0$. The band
inversion happens between the band-2 (eigenenergy $E_{2}$) and the band-3 ($%
E_{3}$) at the critical point specified by Eq. (\ref{QPT}). Hence the rotation eigenvalue
changes by $e^{i2\pi /3}$ at $K$ point. Meanwhile, the rotation eigenvalue
also changes by $e^{i2\pi /3}$ at $K^{\prime}$. The total change of the
rotation eigenvalue by $e^{i4\pi /3}$ indicates that the change of the Chern
number is $-2$ ($+1$ is ruled out because the band inversion occurs at two points: both at $K$ and $K^{\prime}$ points).   On the other hand, for the case of $M<0$, the band inversion happens between the band-1 and the band-3, reflecting in the change of the Chern number by $2$. A consistent explanation has been given by referring to Fig.~\ref{fig:PD_FM}(a). 

\subsection{ $d+id$ superconductor}

\label{sec:theory-classD1}

In this section, we examine the topology of the electronic structures for the
singlet $d+id$ superconducting state. We start by first analyzing a generic
feature of energy spectrum for quasi-particles in the singlet
superconductors with spin $S_z$ conservation. The general BdG Hamiltonian for quasi-particles for singlet
superconductivity can be written as
\begin{equation}
\hat{H}=\frac{1}{2}\sum\limits_{\mathbf{k}}\phi _{\mathbf{k}}^{\dag }%
\mathcal{H}_{\mathbf{k}}\phi _{\mathbf{k}},
\end{equation}%
where $\phi _{\mathbf{k}}= (
\begin{array}{cccc}
c_{\mathbf{k}\uparrow } & c_{\mathbf{k}\downarrow } & c_{-\mathbf{k}\uparrow
}^{\dag } & c_{-\mathbf{k}\downarrow }^{\dag }%
\end{array}
) ^{T}$ with $c_{\mathbf{k}\sigma }$($c_{\mathbf{k}\sigma }^{\dag }$) being
possibly a multi-component vector by including orbital degrees of freedom.
For singlet superconductivity with $S_{z}$ being conserved, the Bloch Hamiltonian is generally given by
\begin{equation} \label{gH}
\mathcal{H}_{\mathbf{k}}=\left(
\begin{array}{cccc}
\xi _{\mathbf{k}\uparrow } & 0 & 0 & \Delta _{\mathbf{k}} \\
0 & \xi _{\mathbf{k}\downarrow } & -\Delta _{\mathbf{k}} & 0 \\
0 & -\Delta _{\mathbf{k}}^{\dag } & -\xi _{-\mathbf{k}\uparrow }^{T} & 0 \\
\Delta _{\mathbf{k}}^{\dag } & 0 & 0 & -\xi _{-\mathbf{k}\downarrow }^{T}
\end{array}%
\right) ,
\end{equation}%
where $\xi _{\mathbf{k}\sigma }$ and $\Delta _{\mathbf{k}}$ can be numbers
or matrices and $\Delta_{\mathbf{-k}} =\Delta_{\mathbf{k}}$. It is clear that the Hamiltonian is block-diagonal and can be
decomposed into two sub-Hamiltonians with $\sigma_z = \pm 1$ characterizing
quasi-particles in each block
\begin{eqnarray}
\hat{H} &=&\frac{1}{2}\sum\limits_{\mathbf{k}}\left\{ \phi _{\uparrow
\mathbf{k}}^{\dag }\left(
\begin{array}{cc}
\xi _{\mathbf{k}\uparrow } & \Delta _{\mathbf{k}} \\
\Delta _{\mathbf{k}}^{\dag } & -\xi _{-\mathbf{k}\downarrow }^{T}%
\end{array}%
\right) \phi _{\uparrow \mathbf{k}}\right. \\
&&\left. +\phi _{\downarrow \mathbf{k}}^{\dag }\left(
\begin{array}{cc}
\xi _{\mathbf{k}\downarrow} & -\Delta _{\mathbf{k}} \\
-\Delta _{\mathbf{k}}^{\dag } & -\xi _{-\mathbf{k}\uparrow }^{T}%
\end{array}%
\right) \phi _{\downarrow \mathbf{k}}\right\},  \notag
\end{eqnarray}
where $\phi _{\uparrow \mathbf{k}} = (
\begin{array}{cc}
c_{\mathbf{k}\uparrow } & c_{-\mathbf{k}\downarrow }^{\dag}%
\end{array}
) $ and $\phi _{\downarrow \mathbf{k}} = (
\begin{array}{cc}
c_{\mathbf{k}\downarrow } & c_{-\mathbf{k}\uparrow }^{\dag}%
\end{array}
)$ are quasi-particle operators for spin up and down respectively. Except
for a minus sign in the paring amplitude or the pairing matrix $\Delta _{
\mathbf{k}}$, Hamiltonians for quasi-particles of both spins are the same.
Hence the eigenenergies $E_{\mathbf{k}}$ of quasi-particles are degenerate
in spin space (the minus sign can be made to be positive by a rotation in
the space of the sub-Hamiltonian with respect to $z$ axis), reflecting the
U(1) spin rotation symmetry. Under the particle-hole transformation, $\phi
_{\uparrow \mathbf{k}} \rightarrow \phi _{\uparrow -\mathbf{k}}^{\dag}$, it
switches two sub-Hamiltonians. The global superconducting state in Eq.(\ref{gH}) thus has PHS. If $
\Delta _{\mathbf{k}}^{\dag } \neq \Delta _{\mathbf{k}}$, there is no TRS.
However, since $S_z$ is conserved, the classification of the topology is based on each sub-Hamiltonian.
\cite{Schnyder2008}
The resulting superconducting state is a combination of two sub-systems in class A with TRS, PHS and chiral being absent in each sub-Hamiltonian. Note that each sub-Hamiltonian
is characterized by pseudo-spin $\mathbf{\tau}$ in the charge sector and in
general, $\xi _{\mathbf{k} \sigma} \neq \xi _{\mathbf{k} -\sigma}$,
there is no PHS symmetry within each sub-Hamiltonian. Since two
sub-Hamiltonians are related by the particle-hole transformation, the
topological invariants for both spin components are the same. Hence the
topological invariant of the whole system is twice of the topological
invariant of any subsystem, indicating a $2\mathbb{Z}$ superconductor. The non-trivial topological state will provide Majorana edge modes.

In the following, we will illustrate that there is a QAH
state in the superconducting state. Mathematically, we find that in each
spin space, there exists a dual transformation that maps the superconducting
state into a ferromagnetic insulating state. We consider the
$d+id$ superconductivity in the tight-binding model of graphene
\begin{eqnarray}
\hat{H}_{\mathrm{SC}} &=&-t\sum_{\left\langle i,j\right\rangle }c_{i}^{\dag
}c_{j}+i\frac{\lambda _{\mathrm{SO}}}{3\sqrt{3}}\sum_{\left\langle
\left\langle i,j\right\rangle \right\rangle }\nu _{ij}c_{i}^{\dag }\sigma
_{z}c_{j}  \label{Eq14} \\
&&+\frac{1}{2}\sum_{\left\langle i,j\right\rangle }\left[ \Delta
_{ij}c_{i}^{\dagger }(i\sigma _{y})\left( c_{j}^{\dagger }\right) ^{T}+%
\mathrm{H.c.}\right] -\mu \sum_{i}c_{i}^{\dag }c_{i}.  \notag
\end{eqnarray}%
Here for the $d+id$ pairing, we have $\Delta _{i+\mathbf{d}_{l},i}=-\frac{2}{%
3}\Delta e^{-i\theta _{l}}$ ($l=1,2,3$; $\theta _{l}$ increases
counterclockwise). After the Fourier transformation, in the basis $c_{%
\mathrm{SC,\uparrow }}(\mathbf{k})=(%
\begin{array}{cccc}
c_{A\mathbf{k}\uparrow } & c_{B\mathbf{k}\uparrow } & -c_{A-\mathbf{k}%
\downarrow }^{\dag } & c_{B-\mathbf{k}\downarrow }^{\dag }%
\end{array}%
)^{T}$, we find that the Bloch Hamiltonian for spin up component is given by
\begin{equation}
\mathcal{H}_{\mathrm{SC,\uparrow }}(\mathbf{k})=\left(
\begin{array}{cccc}
\Lambda _{\mathbf{k}}-\mu  & T_{\mathbf{k}} & 0 & \Delta _{\mathbf{k}} \\
T_{\mathbf{k}}^{\ast } & -\Lambda _{\mathbf{k}}-\mu  & -\Delta _{-\mathbf{k}}
& 0 \\
0 & -\Delta _{-\mathbf{k}}^{\ast } & -\Lambda _{\mathbf{k}}+\mu  & T_{%
\mathbf{k}} \\
\Delta _{\mathbf{k}}^{\ast } & 0 & T_{\mathbf{k}}^{\ast } & \Lambda _{%
\mathbf{k}}+\mu
\end{array}%
\right) .  \label{HSC}
\end{equation}%
Here $T_{\mathbf{k}}$ and $\Lambda _{\mathbf{k}}$ are defined as before. $%
\Delta _{\mathbf{k}}$ has the same form as $R_{\mathbf{k}}$ except that $%
\lambda _{\mathrm{R}}$ is replaced by $\Delta $. Comparing Eqs. (\ref{HSC})
and (\ref{HFM}), we see that the $d+id$ superconductor and the ferromagnetic
insulator have equivalent mathematical structure with the following dual
transformation: \newline
\begin{table}[th]
\centering
\begin{tabular}{ccc}
$d+id$ superconductor &  & ferromagnetic insulator \\[0.5ex] \hline
$\Delta$ & $\leftrightarrow$ & $\lambda_{\mathrm{R}}$ \\
$\mu$ & $\leftrightarrow$ & $M$ \\
\end{tabular}
\label{table1}
\end{table}

\noindent Clearly, $\Delta$ and $\mu$ in $d+id$ superconductors are dual to $\lambda_{\mathrm{R}}$ and $M$ in ferromagnetic insulators. However, the
mechanism for breaking TRS and PHS is different: TRS is
broken by the complex pairing potential in Eq. (\ref{HSC}) but it is broken
by the exchange potential in Eq. (\ref{HFM}) and $\Delta$ and $M$ are not
dual to each other. In addition, PHS is broken by $\lambda _{\mathrm{SO}}$
in Eq. (\ref{HSC}) but it is by $\lambda _{\mathrm{R}}$ in Eq. (\ref{HFM}).
Note that PHS is broken in each sub-Hamiltonian but it is restored in the
superconducting state by including the spin down component.

The existence of the dual transformation implies that the topological
invariants for Eqs. (\ref{HSC}) and (\ref{HFM}) are the same. Therefore, the
topological phases with non-vanishing Chern numbers are the same as in Fig. \ref{fig:PD_FM}(a) by replacing $M$ by $\mu $ and $\lambda _{\mathrm{R}}$ by $%
\Delta $ and doubling the Chern numbers, shown in Fig. \ref{fig:PD_FM}(b).  It further implies that these
phases with non-vanishing Chern numbers become chiral superconducting phases. Specifically, according to the bulk-edge correspondence for Eq. (%
\ref{HFM}), if the Chern number is $C$, one gets $C$ edge states at one edge
and the other $C$ edge states at the opposite edge in a ribbon. In superconductors, particles (with momentum $\mathbf{k}$) and holes ($-\mathbf{k}$) are mixed, and consequently, a particle-like (positive energy) edge mode will show accompanying with a hole-like (negative energy) mode. Here for class D superconductors, these edge modes are
Majorana fermions and thus these particle-like and hole-like modes are not independent but self-charge-conjugate as a Majorana fermion. Hence
the Hamiltonian for corresponding edge states can be generally written as
\begin{eqnarray} \label{edge1}
\hat{H}_{\mathrm{edge}} &=& \sum_{p } \sum^{C}_{i=1} \left( E_p \gamma_{p, i}^\dagger
\gamma_{p, i} + \bar{E}_p \bar{\gamma}_{-p, i}^\dagger \bar{\gamma}%
_{-p, i} \right)  \\
&=& \sum_{p} \sum^{C}_{i=1} \left( E_p \gamma_{-p, i}
\gamma_{p, i} - \bar{E}_p \bar{\gamma}_{-p, i} \bar{\gamma}%
_{p, i} \right), \notag
\end{eqnarray}
where $\gamma_{p, i}$ and $\bar{\gamma}_{p, i}$ are
the Majorana fermion operators at one and the other edges with
corresponding energies being $E_p$ and $\bar{E}_p$, respectively. In above, we only consider positive modes, $E_p, \bar{E}_p \geq 0$. Majorana fermions satisfy $\gamma_{p, i} ^{\dag }=\gamma_{-p, i}$ and $\left\{ \gamma_{-p, i},\gamma_{p^\prime , j} \right\} = \delta_{ij} \delta_{p p^\prime }$. $C$ is
the Chern number and $p$ is the momentum along the edge. These edge modes are chiral so that $E_{-p} = -E_p$.
When there exists, or roughly, in-plane inversion symmetry in the ribbon, we obtain $\bar{E}_p=E_p$ and edge modes at two edges propagating in opposite directions.

For the Hamiltonian $\mathcal{H}_{\mathrm{SC,\uparrow }}(\mathbf{k})$, the Chern number of the quasi-HSC phase 
that corresponds to the SCI/QSH phase also vanishes. For $d+id$ superconductors, 
this only implies $C_{\uparrow }=0$. Since $\mathcal{H
}_{\mathrm{SC,\downarrow }}(\mathbf{k})$ has the same topology, we find that
$C_{\downarrow }=0$. Therefore, the spin Chern number $C_{s}$, which is $%
(C_{\uparrow }-C_{\downarrow })/2$, vanishes. However, the quasi-HSC
phase does still carry a topological invariant. Analogous to the spin Chern number in a QSH phase, the topological invariant is the pseudo-spin Chern number in the charge space of $%
\mathcal{H}_{\mathrm{SC,\uparrow }}(\mathbf{k})$. If one defines the
pseudo-spin quantum number $\tau $ such that $c_{A\mathbf{k}\uparrow }$ and $%
c_{B\mathbf{k}\uparrow }$ have the quantum number $\tau =1$, while
pseudo-spin quantum numbers for $-c_{A-\mathbf{k}\downarrow }^{\dag }$ and $%
c_{B-\mathbf{k}\downarrow }^{\dag }$ are $\tau =-1$, the pseudo-spin Chern number
\begin{equation}
C_{q}=\frac{C_{\tau =1}-C_{\tau =-1}}{2},
\end{equation}%
where $C_{\tau}$ is the Chern number in the $\tau$ sector. $C_{q}$ is one for both $\mathcal{H}_{\mathrm{SC,\uparrow }}(\mathbf{k})$
and $\mathcal{H}_{\mathrm{SC,\downarrow }}(\mathbf{k})$, so the
quasi-HSC phase in the $d+id$ superconductor is characterized by $C_{q}=2$. Since the Chern number
vanishes in this phase, we obtain quasi-helical edge states with similar
Hamiltonian being given by Eq. (\ref{edge1}) with $C=1$ and both $\gamma
_{p,i}$ and $\bar{\gamma}_{p,i}$ describe particle-like
quasi-particles at the same edge.

The phase diagram of topology for the $d+id$ superconductor can be also
understood from the local pairing symmetry in the momentum space. When the
chemical potential $\mu $ lies within the gap, $\lambda _{ \mathrm{SO}}$, of
normal states, since $\lambda _{ \mathrm{SO}} >\Delta$, the system is more
like a band insulator than a superconductor and therefore it behaves like a
QSH insulator. Note that in reality, superconductivity in an insulator can
be induced through proximity effect. As $\mu $ is tuned to go beyond the gap
and cut the band, the gap of the system is determined by the pairing
potential. By expressing the pairing term in the energy basis of electrons
in the normal states, we find
\begin{equation}
c_{A\mathbf{k}\uparrow }c_{B-\mathbf{k} \downarrow }\sim \frac{T_{\mathbf{k}%
}^{\ast }}{2\lambda _{\mathrm{SO}}}\left( c_{+,\mathbf{k}\uparrow } c_{+,-%
\mathbf{k}\downarrow }-c_{-,\mathbf{k}\uparrow } c_{-,\mathbf{k}\downarrow
}\right),
\end{equation}
where $c_{\pm }$ stand for the upper (lower) energy band near the chemical
potential and $T_{K/K^{\prime }+\mathbf{q}}^{\ast }\sim t(q_{x}\pm iq_{y})$ for $%
q\ll \pi $. Clearly, if the pairing amplitude is $\Delta_{\mathbf{k}}$, the
effective pairing symmetry becomes
\begin{equation}
\Delta_{eff} (\mathbf{k}) \sim \Delta_{\mathbf{k}} T^{\ast}_{\mathbf{k}}.
\end{equation}
It is therefore clear that when the pairing function is isotropic \textit{s}%
-wave, the effective pairing symmetry is $p\pm ip$-wave superconductivity
locally at Dirac points. However, due to opposite Berry curvatures at $K$
and $K^{\prime}$, the Chern number vanishes in total for \textit{s}-wave.
For $d+id$ pairing, however, TRS is broken so that
local gap functions at $K$ and $K^{\prime}$ are not equivalent, which
results in nontrivial topology. By performing local expansion near Dirac
points, we find that $\Delta _{K+\mathbf{q}}\sim \Delta (q_{x}+iq_{y})$,
while $\Delta _{K^{\prime}+\mathbf{q}}\sim \Delta $. Therefore, local Berry
curvatures at $K$ and $K^{\prime}$ do not get canceled and both $c_{\pm }$
bands get non-vanishing finite Chern numbers.

Finally, we note in passing that while in the above dual transformation,
only NN pairing and the Rashba SOC are
considered, the duality is valid when NNN couplings are included. Specifically, for the NNN pairing on the same
sub-lattice, there is also a corresponding dual SOC term in the ferromagnetic insulating system. For $d+id$-wave superconductors, we find that the dual term to the
NNN pairing order parameter is the NNN Rashba spin-orbit interaction that is
shown to exist in silicene due to the buckled structure. \cite%
{Liu2011,Ezawa2012b}

\subsection{Class D superconductor with ferromagnetism}

\label{sec:theory-classD2}

We now combine both ferromagnetic and superconducting models in Eqs. (\ref%
{HFM0}) and (\ref{Eq14}). This would be the model to describe the situation
that occurs when a $d+id$ superconductor is placed in proximity to a
ferromagnet so that the exchange field is induced in the $d+id$
superconducting state. In this case, both the Rashba spin-orbit interaction and the pairing potential
are present, hence $S_z$ is no longer conserved. The system belongs to
the class D superconductor.\cite{Schnyder2008}

If we adopt the basis for the electron, $\psi =%
\dbinom{c_{\mathrm{FM}}(\mathbf{k})}{c_{\mathrm{FM}}^{\dag }(-\mathbf{k})}$
where $c_{\mathrm{FM}}(\mathbf{k})$ is the same basis used in Eq. (\ref{HFM}), the
Bloch Hamiltonian is given by
\begin{equation}
\mathcal{H}_{\mathrm{SC/FM}}(\mathbf{k})=\left(
\begin{array}{cc}
\mathcal{H}_{\mathrm{FM}}(\mathbf{k})-\mu \mathbb{I} & \mathcal{D}(\mathbf{k}%
) \\
\mathcal{D}^{\dag }(\mathbf{k}) & -\mathcal{H}_{\mathrm{FM}}^{T}(-\mathbf{k}%
)+\mu \mathbb{I}%
\end{array}%
\right) ,  \label{H_FMSC}
\end{equation}%
where $\mathcal{H}_{\mathrm{FM}}(\mathbf{k})$ is given by Eq. (\ref{HFM})
and $\mathcal{D}(\mathbf{k})$ is the pairing matrix given by
\begin{equation}
\mathcal{D}(\mathbf{k})=\left(
\begin{array}{cccc}
0 & 0 & 0 & -\Delta _{\mathbf{k}} \\
0 & 0 & -\Delta _{-\mathbf{k}} & 0 \\
0 & \Delta _{\mathbf{k}} & 0 & 0 \\
\Delta _{-\mathbf{k}} & 0 & 0 & 0%
\end{array}%
\right) .
\end{equation}%
Note that the Hamiltonian can not be decomposed into two sub-Hamiltonians due to nonconserving $S_z$. In addition, although the Rashba SOC breaks inversion symmetry, in the proximity effect,
neglecting the triplet pairing is an acceptable approximation.

The Hamiltonian $\mathcal{H}_{\mathrm{SC/FM}}(\mathbf{k})$ is self-dual. This
can be seen by constructing a unitary transformation $U$ that brings $\psi $
into the form $\dbinom{c_{\mathrm{SC}}(\mathbf{k})}{c_{\mathrm{SC}}^{\dag }(-%
\mathbf{k})}$. Let $\dbinom{c_{\mathrm{SC}}(\mathbf{k})}{c_{\mathrm{SC}%
}^{\dag }(-\mathbf{k})}=U\psi $, we find that $U$ is given by
\begin{equation}
U=\left(
\begin{array}{cccc}
1 & 0 & 0 & 0 \\
0 & 0 & 0 & -\sigma _{z} \\
0 & 1 & 0 & 0 \\
0 & 0 & -\sigma _{z} & 0%
\end{array}%
\right) ,
\end{equation}%
where $1$ and $\sigma _{z}$ are $2\times 2$ matrices. As a consequence of
duality, one finds
\begin{equation}
U\mathcal{H}_{\mathrm{SC/FM}}(M,\mu ,\lambda _{\mathrm{R}},\Delta )U^{\dag }=%
\mathcal{H}_{\mathrm{SC/FM}}(\mu ,M,\Delta ,\lambda _{\mathrm{R}}).
\end{equation}%
Therefore, the topology of electronic structures for $d+id$ superconductors
in proximity to ferromagnets is symmetric between ($M$, $\mu $) and ($%
\lambda _{\mathrm{R}}$, $\Delta $). This implies that investigating weak
superconductivity ($\Delta <\lambda _{\mathrm{R}}$) and weak ferromagnetism (%
$M<\mu $) allows one to access topological phases of strong
superconductivity and strong ferromagnetism. In reality, since it is not
easy to change $\Delta $ and $M$, the duality allows one to
tune $\lambda _{\mathrm{R}}$ and $\mu $ to access different topological phases of the
system.

\begin{figure}[tbp]
\includegraphics[width=0.48\textwidth]{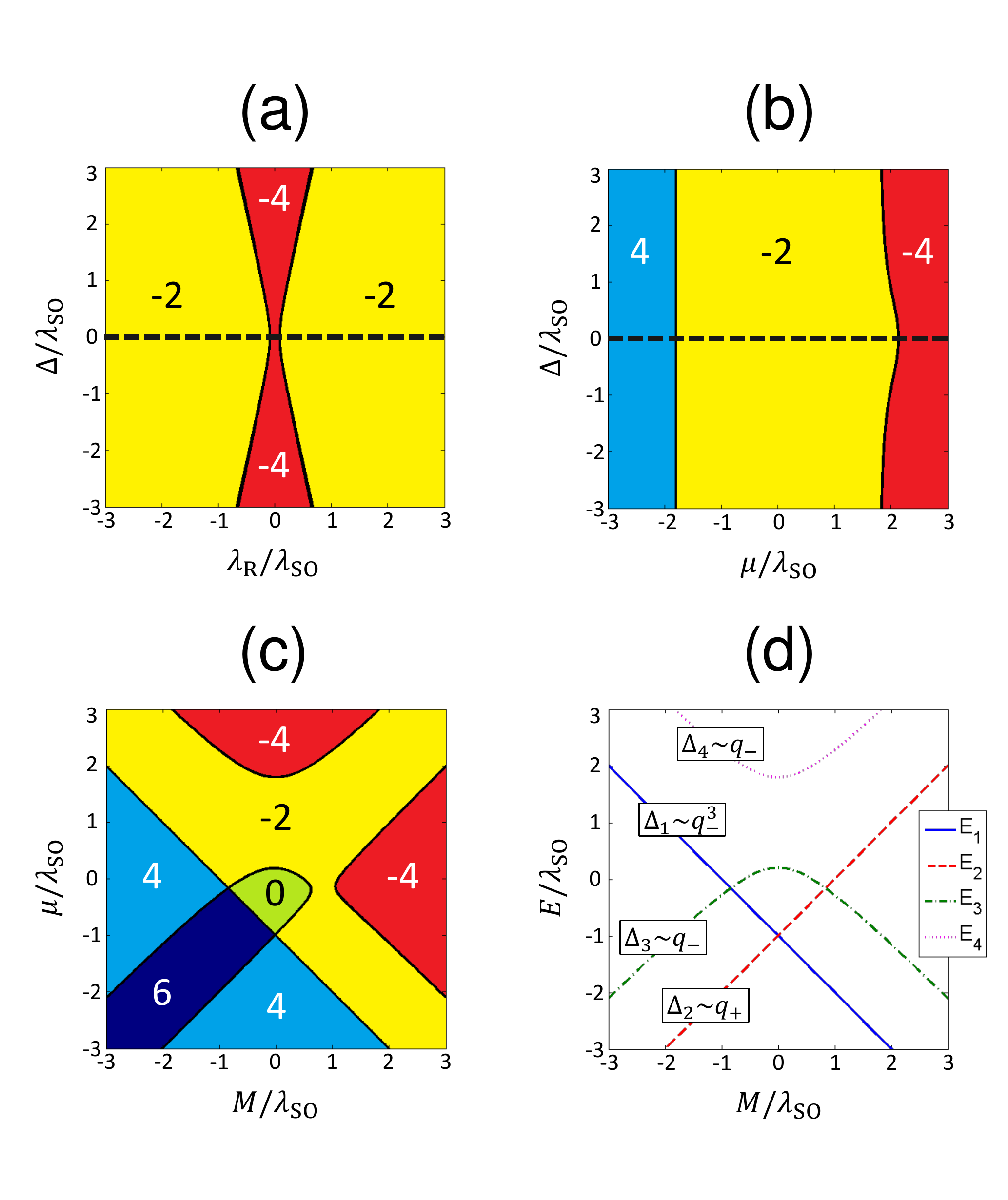}\newline
\caption{(Color online) Typical topological phase diagrams for a $d+id$
superconductor in proximity to a ferromagnet. Numbers shown in colored areas are the corresponding Chern numbers. (a) $\Delta$ versus the Rashba
spin-orbit coupling $\protect\lambda_{\mathrm{R}}$ at $\protect\mu/\lambda _{\mathrm{SO}}=1.2$ and $%
M=0$. (b) $\Delta$ versus $\protect\mu$ for $\protect\lambda_{\mathrm{R}}/\lambda _{\mathrm{SO}}=0.4
$ and $M/\lambda _{\mathrm{SO}}=0.8$. (c) Phase diagram of weak superconductivity ($\Delta /\protect%
\lambda _{\mathrm{SO}}=0.1$) for $\protect\mu$ versus $M$. Here $t/\protect%
\lambda _{ \mathrm{SO}}=10$ and $\protect\lambda _{\mathrm{R}}/\protect%
\lambda _{\mathrm{SO}}=0.4$. (d) Eigenenergies at the $K$ point for four
normal-state bands ($\Delta =0$) as functions of $M$. Here $\protect\lambda %
_{\mathrm{R}}/\lambda _{\mathrm{SO}}=0.4$. Tagged boxes display gap functions near $K$ for
corresponding bands (see text).}
\label{fig:CN_FMSC}
\end{figure}

Typical topological phase diagrams for $\mathcal{H}_{\mathrm{SC/FM}}$ are
shown in Fig. \ref{fig:CN_FMSC}. Here Figure \ref{fig:CN_FMSC} (a) shows
different chiral superconducting phases for a pure $d+id$ superconductor
with $M=0$. It is seen that for small gap amplitude (weak coupling limit), $%
d+id$-wave superconductors are always in chiral superconducting phases with
Chern number being $\pm 2$. Figure \ref{fig:CN_FMSC} (b) shows chiral
superconducting phases for a $d+id$ superconductor mixed with moderate
ferromagnetism. The largest Chern number can go up to 4. In Fig. \ref%
{fig:CN_FMSC} (c), we show the topological phase diagram for weak
superconductivity in $\mu -M$ space at $\left( t/\lambda _{\mathrm{SO}}\text{%
, } \lambda _{\mathrm{R}}/\lambda _{\mathrm{SO}}\text{, }\Delta /\lambda _{
\mathrm{SO}}\right) =\left( 10\text{, }0.4\text{, }0.1\right) $. The Chern
number can be 0, $-2$, $\pm 4$, and $6$ with $6$ being the largest possible
Chern number in this system. The zero Chern number at the center of the
phase diagram is the quantum pseudo-spin Hall phase in the charge sector
with $C_q=2$. As indicated before, the phase boundaries are the band
touching loci and the change of Chern number across the boundaries is the
change of total angular momenta of the filled bands. Because the exchange of
angular momentum happens simultaneously at $K$ and $K^{\prime}$, the Chern
number must be even.

\begin{figure}[tbp]
\includegraphics[width=0.48\textwidth]{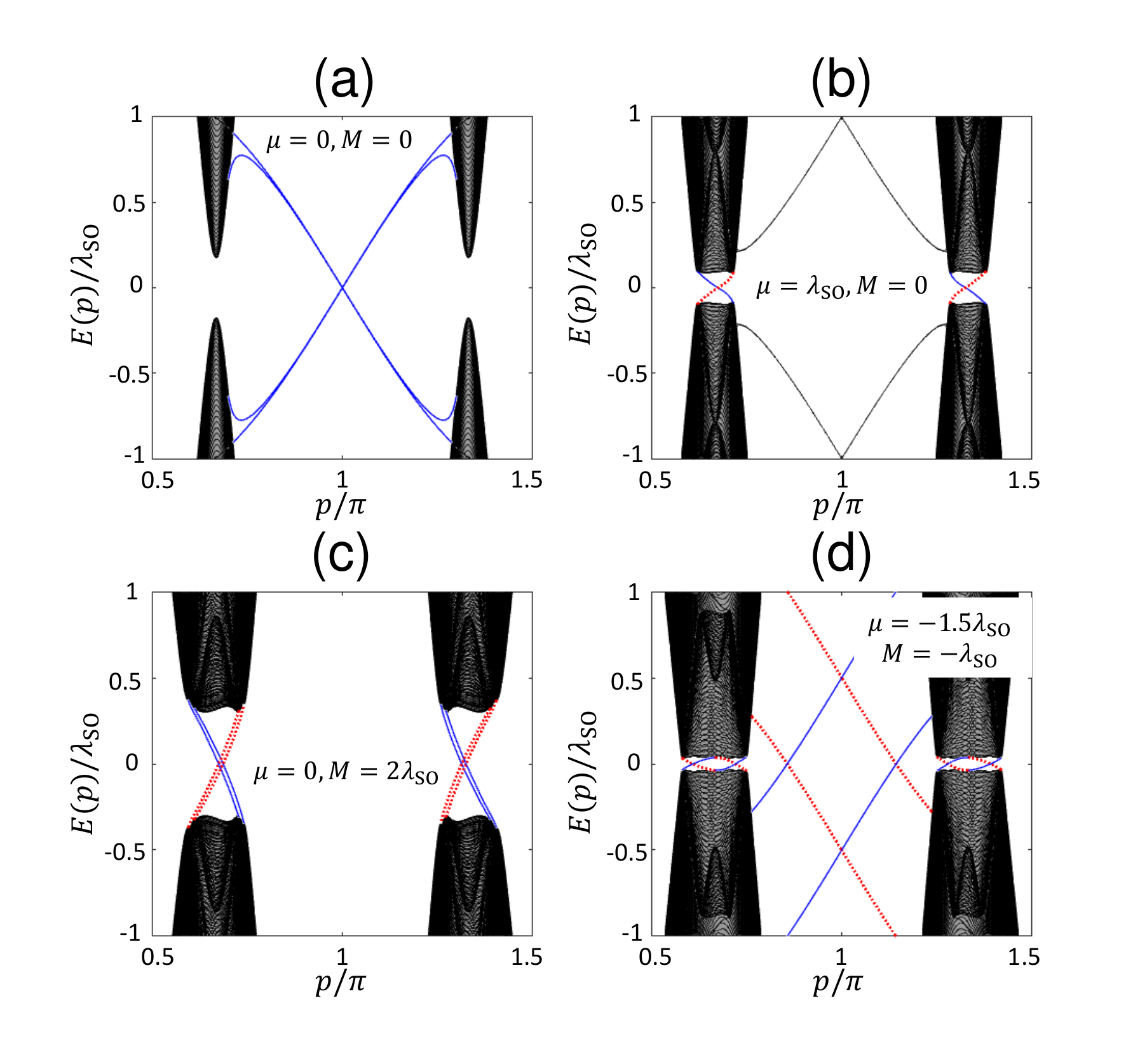}\newline
\caption{(Color online) Bulk-edge correspondence for zigzag ribbons in
different topological superconducting phases. (a) $C=0$, $C_q=2$, (b) $C=-2$ (c) $C=-4$, and
(d) $C=+6$, respectively. Here red and blue lines denote edge states at
opposite edges. The common parameters are $\left( t/\lambda _{\mathrm{SO}}\text{, }\protect\lambda _{%
\mathrm{R}}/\lambda _{\mathrm{SO}}\text{, }\Delta/\lambda _{\mathrm{SO}} \right) =\left( 10\text{, }0.4\text{, }%
0.1\right) $. The number of lattice
points for width of the ribbons is 800.}
\label{fig:zigzag_ribbon}
\end{figure}

Figure \ref{fig:zigzag_ribbon} illustrates the bulk-edge correspondence for
the phase diagram shown in Fig. \ref{fig:CN_FMSC} (c). Here the spectra of a
zigzag ribbon in different topological phases are computed to check the
consistency of equality of the bulk Chern number and the number of chiral
edge modes. Here in the presence of in-plane inversion symmetry for each chiral edge mode at one edge with energy and
momentum $(E,p)$, there is another mode at the other side with $(E,-p)$, and they are Majorana fermions, in agreement with Eq. (\ref{edge1}).

We now illustrate the physical mechanism for the behavior of phase diagram
in Fig. \ref{fig:CN_FMSC}(c) in the weak coupling limit. In this limit, the
pairing is weak and hence considering pairing between intraband electrons is
sufficient. A superconducting state can be generally described by
\begin{equation}
\mathcal{H}(q)=\left(
\begin{array}{cc}
\xi _{q} & \Delta _{q} \\
\Delta _{q}^{\ast } & -\xi _{-q}%
\end{array}%
\right) .  \label{H_LE}
\end{equation}%
Here $q$ is the momentum relative to $K$ or $K^{\prime }$ point, $\xi _{q}$
is the kinetic energy relative to the chemical potential, and $\Delta _{q}$
is the pairing gap function. For a nontrivial superconducting state, the
chemical potential is within an energy band so that $\xi _{q}$ must change
sign in the Brillouin zone. Furthermore, the gap function is generally a
representation of a rotational group and can be generally expressed as $%
\Delta _{q}=\Delta q_{\pm }^{n}$ with $q_{\pm }=q_{x}\pm iq_{y}$ and $n\in
\mathbb{N}$. As an example, the case of $n=1$ is the well-known $p\pm ip$
CSC. Under a rotation by $\phi $, the gap function is an
eigenfunction to the rotation, $q_{\pm }^{n}\rightarrow e^{\mp in\phi
}q_{\pm }^{n}$. Hence the gap function is generally a representation of
rotational group. In other words, the gap function carries angular momentum.
Now from Eqs. (\ref{rot_K}) and (\ref{rot_mK}), for three-fold rotation $%
\phi =2\pi /3$, the rotation eigenvalues of a Cooper pair for energy bands
1, 2, 3, and 4 are $e^{i4\pi /3}$, $e^{-i4\pi /3}$, $1$, and $1$,
respectively. On the other hand, the $d+id$ pairing carries an extra angular
momentum $l=-1$ ($\eta =e^{i2\pi /3}$) such that the gap functions of bands
have rotation eigenvalues $e^{i6\pi /3}$, $e^{-i2\pi /3}$, $e^{i2\pi /3}$,
and $e^{i2\pi /3}$, respectively. As a result, the gap functions behave as $%
\Delta _{1}\sim q_{-}^{3}$, $\Delta _{2}\sim q_{+}$, $\Delta _{3}\sim q_{-}$%
, and $\Delta _{4}\sim q_{-}$. Detailed derivations for the rotation
symmetry in superconducting states are referred to Appendix \ref%
{sec:app-rotation_sc}.

In Fig. \ref{fig:CN_FMSC}(d), we show eigenenergies of four bands $%
E_{1,2,3,4}$ and their corresponding gap functions $\Delta _{1,2,3,4}$.
Comparing Figs. \ref{fig:CN_FMSC}(c) and (d), the close similarity between
the phase boundaries and eigenenergies are found with a small discrepancy
that might result from finite $\Delta /\lambda _{\mathrm{SO}}=0.1$. Since
the eigenenergies at $K$ (or $K^{\prime }$) are the top or the
bottom of the associated energy bands, when the chemical potential passes
across these energies, it can be either falling within an energy band or
outside an energy band, depending on the direction of the chemical potential
moves. Clearly, this indicates a topological phase transition and hence a
change in the Chern number. The value of the changed Chern number is
determined by the gap functions and the direction of chemical potential
enters/leaves a band. By including both contributions from $K$ and $%
K^{\prime }$, the topological phase transitions in Fig. \ref{fig:CN_FMSC}(c)
are reproduced by using Fig. \ref{fig:CN_FMSC}(d). We refer readers to
Appendix \ref{sec:app-CN} for more detailed explanations.

\section{Discussion and summary}

In summary, we have demonstrated a dual transformation between a $d+id$ superconductor and a ferromagnetic insulator in a honeycomb lattice with the former being a CSC and the latter being a QAH insulator. The $d+id$ superconductor can be viewed as a combination of two QAH insulators, which map into one another under particle-hole transformation, and thus carries a $2\mathbb{Z}$ topological invariant. When the superconducting pairing amplitude is weak and the chemical potential falls within the SOC gap, the superconductor is quasi-helical and its dual phase is the SCI state with a small Rashba SOC and weak ferromagnetism. Moreover, when both superconductivity and ferromagnetism are included, the system in class D possesses self-duality: $\mu \leftrightarrow M$ and $ \Delta \leftrightarrow \lambda _{\mathrm{R}}$. This implies that topological effects of strong superconductivity or ferromagnetism can be observed in topological states with weak superconductivity or ferromagnetism.

We have also explored the topological superconductor from the effective
low-energy Hamiltonian, Eq. (\ref{H_LE}), in which the
topology is encoded in the gap function $\Delta _{q}$ when the normal-state
Fermi surface is present. The nontrivial topology is present when the gap
function $\Delta _{q}=\Delta q_{+}^{n_{+}}q_{-}^{n_{-}}$ with $%
n_{+}-n_{-}\neq 0$, which can be determined by the rotational eigenvalues: $\Delta
_{\mathfrak{R}\mathbf{q}}=\eta ^{K}\eta ^{K^{\prime }}e^{i\phi }\Delta _{%
\mathbf{q}}$, where $\eta ^{K/K^{\prime }}$ is the phase gained under
rotation from the band electron at $K/K^{\prime }$ and $e^{i\phi }$ from the
$d+id$-wave nature. Equivalently, the criterion for non-trivial topology is to require $\eta ^{K}\eta ^{K^{\prime }}e^{i\phi }\neq 1$. From this inequality, it is clear that
the time-reversal invariant would demand $\eta ^{K}=( \eta
^{K^{\prime }} ) ^{\ast }$ and $e^{i\phi }=1$, hence for a nontrivial two-dimensional superconductor with non-vanishing Chern number, breaking
TRS is essential. The inequality also explains that why an $s$-wave ($e^{i\phi }=1$)
superconductor can be topologically nontrivial if TRS is
broken to in normal states such that $\eta ^{K} \neq ( \eta
^{K^{\prime }} ) ^{\ast }$.

Finally, we discuss experimental features that can be observed for topological superconductors. According to the bulk-edge correspondence, the Chern number for topological superconductors is the number of Majorana edge modes. Since these midgap modes are localized at edges, they will play important roles at low voltage of the tunneling conductance.~\cite{Kao2015, Mou2003} For the Hall conductivity measurements, it is known that the Hall conductivity will be quantized in an insulator as $\sigma_{\mathrm{H}} = C e^2/h $, $C$ the Chern number,~\cite{TKNN} and deviate from the quantized value when doped into a metal. In superconductors, because charges are not conserved,the Hall conductivity is no longer quantized even if the Chern number is non-vanishing. In the weak pairing limit, the change in the Hall conductivity comes from the change in the occupation number and the change of Berry curvature,~\cite{Chung2014} and thus the Hall conductivity decreases against the superconducting gap. Although the Hall conductivity does not show a clear signature to differentiate a topological superconductor from a trivial one, its derivatives deliver the signature of topological phase transitions.~\cite{Sacramento2014}

On the other hand, due to the energy conservation, the topological invariant, Chern numbers, can still be revealed in the thermal Hall conductivity. It is known that the thermal Hall conductivity of the topological superconductors in the low temperature limit is given by $\kappa_{xy}=\frac{C}{2}\frac{\pi T}{6}$ with the coefficient to the temperature $T$ being quantized.~\cite{Read2000,Qin2011a,Sumiyoshi2013} Here the appearance of half the Chern number is a
reflection of the half-fermion nature for Majorana modes. Different topological phases shown in Fig.~\ref{fig:CN_FMSC}(c) can be thus identified by measuring the thermal Hall conductivity.

\begin{acknowledgments}
This work was supported by Ministry of
Science and Technology (MoST), Taiwan. SMH would like to thank Hsin Lin for his helpful suggestions and also appreciate NCTS for accommodation support. We also acknowledge support by Academia Sinica Research Program on Nanoscience and Nanotechnology, Taiwan.
\end{acknowledgments}

\appendix{}

\section{Rotation symmetry for non-superconducting states}

\label{sec:app-rotation_normal}

In this Appendix, we examine the rotation symmetry in the
non-superconducting states and will find the rotation eigenvalues of four
bands at $K$ and $K^{\prime }$ valleys for non-superconducting states. In
the honeycomb lattice, there exists three-fold rotation symmetry, satisfying $%
\left[ \hat{R},\hat{H}\right] =0$, where $\hat{R}$ stands for a clockwise
three-fold rotation. For the Bloch Hamiltonian, it reads%
\begin{equation}
\mathcal{RH}(\mathbf{k})\mathcal{R}^{\dag }=\mathcal{H}(\mathfrak{R}\mathbf{k%
}),  \label{H_rot}
\end{equation}%
where $\mathfrak{R}\mathbf{k}$ is the transformed wave vector $\mathbf{k}$
under rotation and $\mathcal{R}$ is the rotation matrix for a given
representation, resulting from $\hat{R}\psi _{\mathbf{k}}^{\dag }\hat{R}%
^{-1}=\psi _{\mathfrak{R}\mathbf{k}}^{\dag }\mathcal{R}$. Since $K$ and $%
K^{\prime }$ are rotation-invariant momenta, which satisfy $\mathfrak{R}K=K-%
\mathbf{G}_{2}$ and $\mathfrak{R}K^{\prime }=K^{\prime }+\mathbf{G}_{2}$
with reciprocal lattice $\mathbf{G}_{2}=2\pi (1,1/\sqrt{3})$, the energy
eigenstates are thus also the rotation eigenstates. After rotation, the
basis will transform as $c_{AK\sigma }^{\dag }\rightarrow c_{A\left( K-%
\mathbf{G}_{2}\right) \sigma }^{\dag }e^{i\sigma \pi /3}=c_{AK\sigma }^{\dag
}e^{i\sigma \pi /3}$ and $c_{BK\sigma }^{\dag }\rightarrow c_{B\left( K-%
\mathbf{G}_{2}\right) \sigma }^{\dag }e^{i\sigma \pi /3}=c_{BK\sigma }^{\dag
}e^{i\sigma \pi /3}e^{-i2\pi /3}$, where $e^{-i2\pi /3}=e^{i\left( -\mathbf{G%
}_{2}\right) \cdot \left( -\mathbf{d}_{1}\right) }$ comes from the
non-primitive sub-lattice vector in the unit cell. By expanding the
Hamiltonian around $K$ and $K^{\prime }$ with $\mathbf{q}$ being the deviation
of momentum, we find
\begin{eqnarray}
\mathcal{RV}_{G}\mathcal{H}^{K}(\mathbf{q})\mathcal{V}_{G}^{\dag }\mathcal{R}%
^{\dag } &=&\mathcal{H}^{K}(\mathfrak{R}\mathbf{q}),  \label{H_eff_K} \\
\mathcal{RV}_{G}^{\dag }\mathcal{H}^{K^{\prime }}(\mathbf{q})\mathcal{V}_{G}%
\mathcal{R}^{\dag } &=&\mathcal{H}^{K^{\prime }}(\mathfrak{R}\mathbf{q}),
\label{H_eff_Kp}
\end{eqnarray}%
where $\mathcal{R}=\exp \left( i\sigma _{z}\pi /3\right) $ and $\mathcal{V}%
_{G}=\exp \left[ i\left( \tau _{z}-1\right) \pi /3\right] $ for $\sigma _{z}$
on spin and $\tau _{z}$ on sub-lattice space. The rotation eigenvalues of
states at $K$ and $K^{\prime }$ are the eigenvalues of $\mathcal{RV}%
_{G}$ and $\mathcal{RV}_{G}^{\dag }$, respectively.

To find the eigenstates, we need the expansion of functions for
the momentum around $K$, $K^{\prime }$ with small deviation $\mathbf{q}$, $T_{K/K^{\prime
}+\mathbf{q}}\approx \pm \frac{\sqrt{3}}{2}tq_{\mp }$, $\Lambda
_{K/K^{\prime }+\mathbf{q}}\approx \mp \lambda _{\mathrm{SO}}$, $R_{K+%
\mathbf{q}}\approx -\frac{i}{\sqrt{3}}\lambda _{\mathrm{R}}q_{+}$ and $%
R_{K^{\prime }+\mathbf{q}}\approx -i2\lambda _{\mathrm{R}}$ with $q_{\pm
}=q_{x}\pm iq_{y}$. The Bloch Hamiltonian at $K$, Eq. (\ref{HFM}), is
\begin{equation}
\mathcal{H}_{\mathrm{FM}}^{K}=\left(
\begin{array}{cccc}
-\lambda _{\mathrm{SO}}-M & 0 & 0 & 0 \\
0 & \lambda _{\mathrm{SO}}-M & i2\lambda _{\mathrm{R}} & 0 \\
0 & -i2\lambda _{\mathrm{R}} & \lambda _{\mathrm{SO}}+M & 0 \\
0 & 0 & 0 & -\lambda _{\mathrm{SO}}+M%
\end{array}%
\right) ,
\end{equation}%
whose eigenstates are given by
\begin{equation}
\begin{array}{l}
\left\vert \gamma _{1}\right\rangle _{K}=\left(
\begin{array}{cccc}
1, & 0, & 0, & 0%
\end{array}%
\right) ^{T}, \\
\left\vert \gamma _{2}\right\rangle _{K}=\left(
\begin{array}{cccc}
0, & 0, & 0, & 1%
\end{array}%
\right) ^{T}, \\
\left\vert \gamma _{3}\right\rangle _{K}=\left(
\begin{array}{cccc}
0, & -i\sin \frac{\theta }{2}, & \cos \frac{\theta }{2}, & 0%
\end{array}%
\right) ^{T}, \\
\left\vert \gamma _{4}\right\rangle _{K}=\left(
\begin{array}{cccc}
0, & \cos \frac{\theta }{2}, & -i\sin \frac{\theta }{2}, & 0%
\end{array}%
\right) ^{T},%
\end{array}
\label{eigfun_K}
\end{equation}%
with the corresponding eigenenergies being
\begin{equation}
\begin{array}{l}
E_{1}=-\lambda _{\mathrm{SO}}-M, \\
E_{2}=-\lambda _{\mathrm{SO}}+M, \\
E_{3}=\lambda _{\mathrm{SO}}-\sqrt{M^{2}+4\lambda _{\mathrm{R}}^{2}}, \\
E_{4}=\lambda _{\mathrm{SO}}+\sqrt{M^{2}+4\lambda _{\mathrm{R}}^{2}},%
\end{array}
\label{eig_energy_K}
\end{equation}%
respectively. Here $\theta =-\arctan \frac{2\lambda _{\mathrm{R}}}{M}$. The
rotation eigenvalues of these four states are obtained by applying $\mathcal{RV}_{G}$. We obtain
\begin{equation}
\begin{array}{l}
\eta _{1}^{K}=e^{i\pi /3}, \\
\eta _{2}^{K}=e^{-i3\pi /3}, \\
\eta _{3}^{K}=e^{-i\pi /3}, \\
\eta _{4}^{K}=e^{-i\pi /3}.
\end{array}
\label{rot_val_K}
\end{equation}%
For the case of $M>0$, the band inversion happens between the
the second $\left\vert \gamma _{2}\right\rangle $ and the third band $%
\left\vert \gamma _{3}\right\rangle $ and hence the rotation eigenvalue
changes by $e^{i2\pi /3}$.

On the other hand, for the $K^{\prime }$ point, the Hamiltonian is
\begin{equation}
\mathcal{H}_{\mathrm{FM}}^{K^{\prime }}=\left(
\begin{array}{cccc}
\lambda _{\mathrm{SO}}-M & 0 & 0 & -i2\lambda _{\mathrm{R}} \\
0 & -\lambda _{\mathrm{SO}}-M & 0 & 0 \\
0 & 0 & -\lambda _{\mathrm{SO}}+M & 0 \\
i2\lambda _{\mathrm{R}} & 0 & 0 & \lambda _{\mathrm{SO}}+M%
\end{array}%
\right) ,
\end{equation}%
and the eigenstates are given by
\begin{equation}
\begin{array}{l}
\left\vert \gamma _{1}\right\rangle _{K^{\prime }}=\left(
\begin{array}{cccc}
0, & 1, & 0, & 0%
\end{array}%
\right) ^{T}, \\
\left\vert \gamma _{2}\right\rangle _{K^{\prime }}=\left(
\begin{array}{cccc}
0, & 0, & 1, & 0%
\end{array}%
\right) ^{T}, \\
\left\vert \gamma _{3}\right\rangle _{K^{\prime }}=\left(
\begin{array}{cccc}
\cos \frac{\theta }{2}, & 0, & 0, & i\sin \frac{\theta }{2}%
\end{array}%
\right) ^{T}, \\
\left\vert \gamma _{4}\right\rangle _{K^{\prime }}=\left(
\begin{array}{cccc}
i\sin \frac{\theta }{2}, & 0, & 0, & \cos \frac{\theta }{2}%
\end{array}%
\right) ^{T}.%
\end{array}
\label{eigfun_Kp}
\end{equation}%
Here the eigenenergies are the same as those given by Eqs.(\ref{eig_energy_K}). From the eigenstates, their rotation
eigenvalues of $\mathcal{RV}_{G}^{\dag }$ are
\begin{equation}
\begin{array}{l}
\eta _{1}^{K^{\prime }}=e^{i3\pi /3}, \\
\eta _{2}^{K^{\prime }}=e^{-i\pi /3}, \\
\eta _{3}^{K^{\prime }}=e^{i\pi /3}, \\
\eta _{4}^{K^{\prime }}=e^{i\pi /3},%
\end{array}
\label{rot_val_Kp}
\end{equation}%
For $M>0$, the band inversion between $\left\vert \gamma _{2}\right\rangle $
and $\left\vert \gamma _{3}\right\rangle $ at $K^{\prime }$ also coincides
with the change of rotation eigenvalue by $e^{i2\pi /3}$.

\section{Rotation symmetry in the superconducting states}

\label{sec:app-rotation_sc}

In this Appendix, we elucidate the rotation symmetry in superconducting
state. We have demonstrated that there is a three-fold rotation symmetry in
the non-superconducting state with the consequence $\left[ \hat{R},\hat{H}%
\right] =0$ in Appendix \ref{sec:app-rotation_normal}. In unconventional
superconductors, the pairing term is not rotation-invariant but it forms a
representation of the rotational group by obeying $\hat{\Delta}\rightarrow
e^{i\phi }\hat{\Delta}$ under a three-fold rotation. Here for the $d+id$%
-wave in our case, $\phi =2\pi /3 $. The extra phase $\phi$ results from
the internal angular momentum
of the Cooper pair and can be combined with the U(1) phase associated with
the spontaneous U(1) gauge symmetry breaking in the superconducting state.
As a result, the rotation symmetry is not really
broken and can be restored by a gauge transformation.

In order to include phases associated with angular momenta of Cooper pairs,
the condition for the rotational symmetry, Eq. (\ref{H_rot}), needs to be modified. If we adopt the same
rotation matrix $\mathcal{R}$ for both particle and hole, a BdG Hamiltonian, ${\mathcal H}_{\mathrm{SC}}(\mathbf{k})$, is rotation-invariant if it satisfies
\begin{equation}
\mathcal{\bar{R}\bar{V}H}_{\mathrm{SC}}(\mathbf{k})\mathcal{\bar{V}^{\dag }\bar{R}}^{\dag }=\mathcal{H}_{\mathrm{SC}}(\mathfrak{R}%
\mathbf{k}),  \label{H_rot2}
\end{equation}%
where $\mathcal{\bar{R}}=\mathrm{diag} (
\begin{array}{cc}
\mathcal{R}, & \mathcal{R}^{\ast }%
\end{array} )$ and $\mathcal{\bar{V}}=\mathrm{diag} (
\begin{array}{cc}
1, & e^{-i\phi }
\end{array} )$ with the first elements ($\mathcal{R}$ and 1) and the second elements ($%
\mathcal{R}^{\ast }$ and $e^{-i\phi }$) acting on particle and hole space,
respectively. The matrix $\bar{V}$ reproduces the phase $\phi$ for superconductors under
the operation of rotation.

We now derive the effective pairing symmetries for bands around $K$ and $K^{\prime }$.
We shall start from the low-energy
Hamiltonian for a given band around $K$ or $K^{\prime }$ by
\begin{equation}
\mathcal{H}_{\mathrm{SC}}^{K/K^{\prime }}(\mathbf{q})=\left(
\begin{array}{cc}
\xi _{\mathbf{q}} & \Delta _{\mathbf{q}} \\
\Delta _{\mathbf{q}}^{\ast } & -\xi _{-\mathbf{q}}%
\end{array}%
\right) .  \label{H_K}
\end{equation}%
Here in the weak-coupling limit, only intra-band pairing is considered and $\Delta _{\mathbf{q}}$ is non-vanishing only in a small energy range around the chemical potential.
Similar to Eqs. (\ref{H_eff_K}) and (\ref{H_eff_Kp}), the rotation symmetry
requires
\begin{equation}
\mathcal{\bar{R}\bar{V}H}_{\mathrm{SC}}^{K/K^{\prime }}(\mathbf{q})%
\mathcal{\bar{V}^{\dag }\bar{R}}^{\dag }=\mathcal{H}_{\mathrm{SC}%
}^{K/K^{\prime }}(\mathfrak{R}\mathbf{q}).
\end{equation}%
For each band, $\mathcal{R}$ is replaced by the corresponding rotation eigenvalues $\eta
^{K/K^{\prime }}$ that are obtained for the non-superconducting state
in Eqs. (\ref{rot_val_K}) and (\ref{rot_val_Kp}) and we obtain
\begin{equation}
\Delta _{\mathfrak{R}\mathbf{q}}=\eta ^{K}\eta ^{K^{\prime }}e^{i\phi
}\Delta _{\mathbf{q}}.  \label{Delta_q}
\end{equation}%
The phase in Eq. (\ref{Delta_q}) results from the rotation of a Cooper pair and consists of
phases from their composite electrons at $K$ and $K^{\prime }$ and the nontrivial phase from the symmetry of the gap function. For the $d+id$%
-wave with $e^{i\phi }=e^{i2\pi /3}$, the gap function can be considered to
carry angular momentum $l=-1$. In general,
the pairing potential near $K$ and $K^{\prime}$ can be expressed as $\Delta _{\mathbf{q}}=\Delta
q_{+}^{n_{+}}q_{-}^{n_{-}}$ with $q_{\pm }=q_{x}\pm i q_{y}$ and $n_{\pm
}\in \mathbb{N}$. By using Eq. (\ref{Delta_q}) and the identity, $\mathfrak{R}q_{\pm }=e^{\mp i2\pi /3}q_{\pm }$,
we find that the total phase associated with $\Delta _{\mathbf{q}}$ is
$\eta ^{K}\eta ^{K^{\prime }}e^{i\frac{2\pi }{3}} =e^{i\frac{2\pi }{3}\left( n_{-}-n_{+}\right)
}$ with $n_{\pm }$ being
determined by $\eta ^{K}$ and $\eta ^{K^{\prime }}$ of the corresponding band. Using $\eta ^{K}$ and $\eta ^{K^{\prime }}$ in Eqs. (\ref
{rot_val_K}) and (\ref{rot_val_Kp}), we conclude that the effective gap functions for four
bands behave as
\begin{equation}
\begin{array}{l}
\Delta _{1}\sim q_{-}^{3}, \\
\Delta _{2}\sim q_{+}, \\
\Delta _{3}\sim q_{-}, \\
\Delta _{4}\sim q_{-}.%
\end{array}%
\end{equation}

\section{Chern Number in the weak coupling limit}

\label{sec:app-CN}

In this Appendix, we present a simple way to understand the topological
number of a superconductor in the weak coupling limit. Here in weak coupling
limit, one assumes that the pairing is weak and only pairing between
intra-band electrons is considered. In the weak coupling, the
superconducting gap function on the Fermi surface of a given band is
sufficient to determine the Chern number of superconductivity.

First of all, we choose the convention for the Berry connection as
\begin{equation}
\mathbf{A}_{n}(\mathbf{k})=i\left\langle u_{n}(\mathbf{k})\left\vert \mathbf{%
\nabla }_{\mathbf{k}}\right\vert u_{n}(\mathbf{k})\right\rangle ,
\end{equation}%
where $u_{n}$ is the Bloch wavefunction of band $n$. In
this convention, the Chern number is the integral of Berry curvature $
\mathbf{\nabla }_{\mathbf{k}}\times \mathbf{A}_{n}(\mathbf{k})$ from the
filled bands over the Brillouin zone and can be formulated as
\begin{eqnarray}
C &=&-\int_{\mathrm{BZ}}\frac{d^{2}k}{2\pi }\sum_{\alpha ,\beta } \\
&& 2 \mathrm{Im} \left\{ \frac{\left\langle u_{\alpha }(\mathbf{k})\left\vert v_{x}(\mathbf{k}
)\right\vert u_{\beta }(\mathbf{k})\right\rangle \left\langle u_{\beta }(
\mathbf{k})\left\vert v_{y}(\mathbf{k})\right\vert u_{\alpha }(\mathbf{k}
)\right\rangle }{\left[ E_{\alpha }(\mathbf{k})-E_{\beta }(\mathbf{k})\right]
^{2}} \right\},  \notag
\end{eqnarray}%
where $\alpha $ ($\beta $) denotes filled (unfilled) bands and $v_{i}(\mathbf{%
k})=\frac{\partial H(\mathbf{k})}{\partial k_{i}}$ ($i=x$, $y$).

\begin{figure}[tbp]
\includegraphics[width=0.48\textwidth]{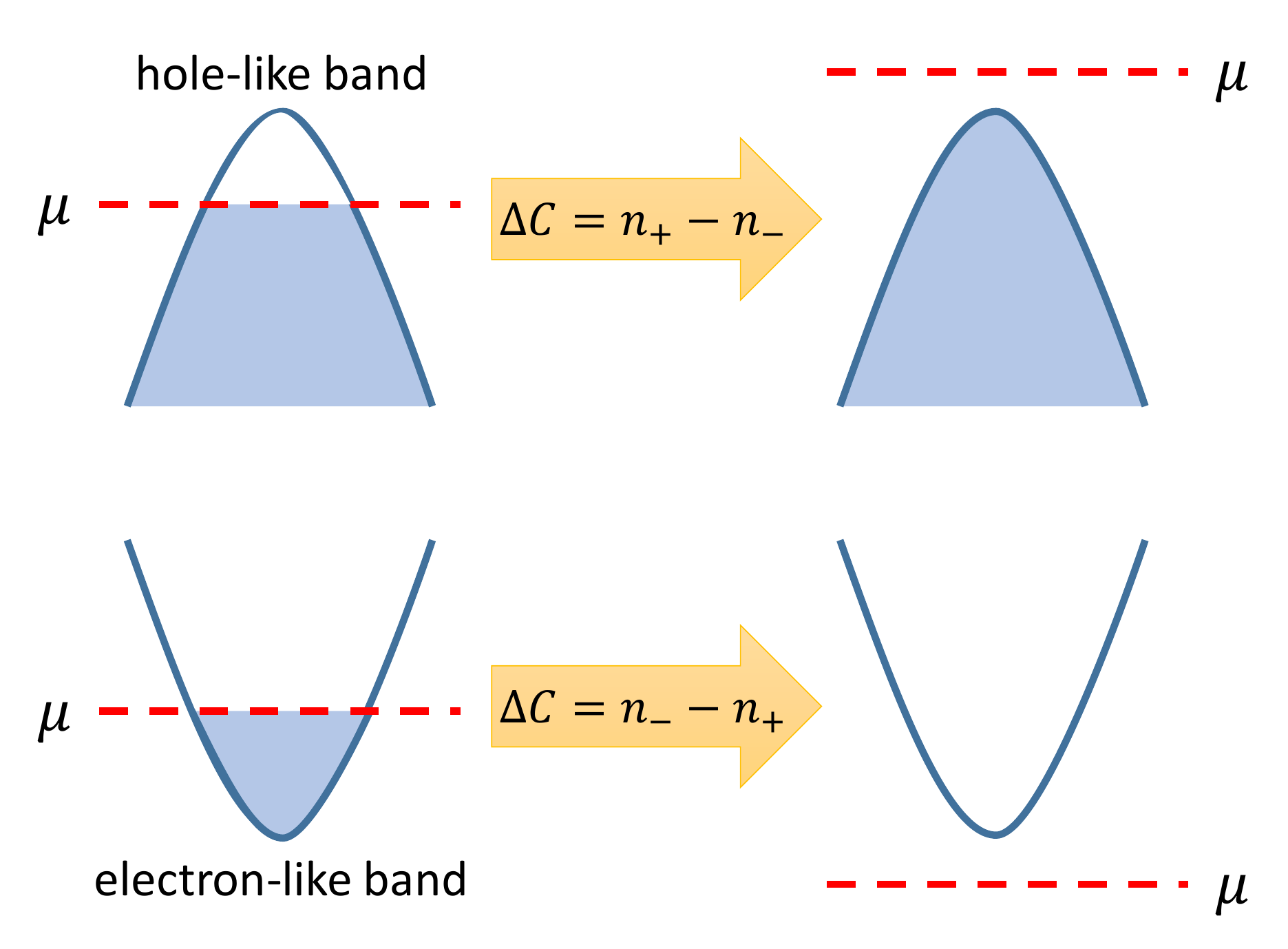}\newline
\caption{(Color online) Change of the Chern number $\Delta C$ of a
superconducting state with a (local) gap function $\Delta
q_{+}^{n_{+}}q_{-}^{n_{-}}$ [refer to Eq. (\protect\ref{Heff})]. Here $%
n_{\pm }\in \mathbb{N}$. A nontrivial superconducting state emerges when the
chemical potential $\protect\mu $ cuts through the band of the normal state.
Upper panel: Change of the Chern number when the normal state goes from hole-like band to be a completely filled band.
Lower panel: Change of the Chern number when the normal state goes from electron-like band to be a unfilled band. Note that the Chern
number is not affected by the sign of the pairing amplitude $\Delta $.}
\label{fig:eh_band}
\end{figure}

Let us start a Hamiltonian in the form
\begin{equation}
\mathcal{H}(q)=\left(
\begin{array}{cc}
\xi _{q} & \Delta q_{+}^{n_{+}}q_{-}^{n_{-}} \\
\Delta ^{\ast }q_{-}^{n_{+}}q_{+}^{n_{-}} & -\xi _{q}%
\end{array}%
\right) ,  \label{Heff}
\end{equation}%
with $q_{\pm }=q_{x}\pm iq_{y}$ and $n_{\pm }\in \mathbb{N}$. The
Hamiltonian describes a superconductor, in which $\xi _{q}$ is the kinetic
energy with respect to the chemical potential and $\Delta
q_{+}^{n_{+}}q_{-}^{n_{-}}$ is the gap function. By using the Pauli matrices $\mathbf{\sigma}$,
the Hamiltonian can be rewritten as
\begin{equation}
\mathcal{H}(q)=E(q)\mathbf{h}_{q}\cdot \mathbf{\sigma } \label{spinform}
\end{equation}%
with $E(q)=\sqrt{\xi _{q}^{2}+\left\vert \Delta \right\vert
q^{2(n_{+}+n_{-})}}$. Here $\mathbf{h}_{q}=(\sin \Theta _{q}\cos \Phi
_{q},\sin \Theta_q \sin \Phi _{q},\cos \Theta _{q})$ is the unit vector with $\Theta _{q}$ and $
\Phi_q$ characterizing its direction. The Chern number can be then expressed as
\begin{eqnarray}
C &=&\frac{1}{4\pi }\int d^{2}q\mathbf{h}_{q}\cdot \frac{\partial \mathbf{h}%
_{q}}{\partial q_{x}}\times \frac{\partial \mathbf{h}_{q}}{\partial q_{y}} \\
&=&\frac{1}{4\pi }\int d^{2}q\epsilon _{ij}\partial _{q_{i}}\cos \Theta
_{q}\partial _{q_{j}}\Phi _{q}.  \notag \label{Ch}
\end{eqnarray}%
Eq.(\ref{spinform}) indicates that the Hamiltonian can be viewed as a mapping from $R^{2}$ in $q$ space to $S^{2}$
in ($\Theta $,$\Phi $) and $\epsilon _{ij}\partial _{q_{i}}\cos \Theta
_{q}\partial _{q_{j}}\Phi _{q}$ is the Jacobian between them such that the
Chern number stands for the covering times of the fields ($\Theta $,$\Phi $)
over a sphere. Therefore, if $\xi $ can change sign with $q$, namely $\xi
_{q=0}\xi _{q\sim \pi }<0$, the chemical potential is within the energy band
and the Chern number is found to be (when the weak coupling condition, $\Delta _{q\sim \pi }=0$, is applied implicitly)
\begin{equation}
C=\mathrm{sgn}(\xi _{q=0})\left( n_{-}-n_{+}\right) .
\end{equation}%
On the other hand, if $\xi $ does not change sign with $q$, the Chern number
is found to be zero. In other words, the Hamiltonian in Eq. (\ref{Heff})
describes a nontrivial superconductor if the chemical potential passes
through the band ($\xi _{q}$ will change sign), otherwise, it is a trivial
superconductor. For a nontrivial superconductor, the normal-state Fermi
surface is hole-like if $\xi _{q=0}>0$ and electron-like if $\xi _{q=0}<0$.
In Fig.~\ref{fig:eh_band}, we illustrates how the Chern number changes when the chemical
potential moves out of the band .
\bibliography{resubmit_ref}
\end{document}